\documentclass[shortnote,twocolumn]{jpsj3}

\usepackage{txfonts}
\usepackage[dvipdfmx]{graphicx}
\usepackage{color}
\usepackage{dcolumn}
\usepackage{bm}
\usepackage{amsmath,amssymb}
\usepackage{ulem}

\begin{document}

\title{Revisiting the phase diagram of T*-type La$_{1-x/2}$Eu$_{1-x/2}$Sr$_x$CuO$_4$ using Oxygen $K$-edge X-ray absorption spectroscopy}

\author{Shun Asano$^{1, 2}$, Kenji Ishii$^3$, Kohei Yamagami$^4$, Jun Miyawaki$^4$, Yoshihisa Harada$^4$, and Masaki Fujita$^2$\thanks{fujita@imr.tohoku.ac.jp}}
\inst{$^1$Department of Physics, Tohoku University, Aoba, Sendai 980-8578, Japan\\
$^2$Institute for Materials Research, Tohoku University, Katahira, Sendai 980-8577, Japan\\
$^3$Synchrotron Radiation Research Center, National Institutes for Quantum and Radiological Science and Technology, Hyogo 679-5148, Japan\\
$^4$Institute for Solid State Physics, The University of Tokyo, Kashiwa 277-8581, Japan} 

\abst{
Oxygen $K$-edge X-ray absorption spectroscopy measurements were conducted on T*-type La$_{1-x/2}$Eu$_{1-x/2}$Sr$_x$CuO$_4$ (LESCO) to estimate the hole density ($n_{\rm h}$) and investigate the oxidation annealing effect on $n_{\rm h}$. A drastic increase in $n_{\rm h}$ due to annealing was found. The increase in $n_{\rm h}$ cannot be explained solely by the oxygen gain due to annealing, suggesting that delocalized holes are introduced into the CuO$_2$ plane. A phase diagram of LESCO was redrawn against $n_{\rm h}$. 
}

\maketitle

In lamellar oxides, the oxygen content and site occupation are essential factors in the emergence of superconductivity. In T'-type $RE_{2-x}$Ce$_x$CuO$_4$ ($RE$ = rare earth), the oxygen reduction due to annealing induces an electron-doping and causes a uniform electrostatic potential on the CuO$_2$ plane by removing chemical disorders such as partially existing apical oxygen (O$_{\rm ap}$) in the as-sintered (AS) sample~\cite{Takagi1989}. It was recently clarified that the number of electrons induced due to annealing process can exceed the expected value from the reduced oxygen content under  the assumption of electron-doping into the upper Hubbard band~\cite{Asano2018}. This result suggests a variation in the band structure due to annealing and the non-unique effects of oxygen reduction on the electronic state. 

Hole-doped T*-type cuprates containing CuO$_5$ pyramids in the unit cell are known to exhibit superconductivity owing to high-pressure oxidation annealing (OA)~\cite{Akimitsu1988}. T*-type AS La$_{1-x/2}$Eu$_{1-x/2}$Sr$_x$CuO$_4$ (LESCO) undergoes a magnetic transition within a wide $x$ range of at least 0.14 - 0.28 and exhibits superconducting (SC) state due to annealing~\cite{Asano2019}. The essential role of annealing in the emergence of superconductivity is considered to be the repair of oxygen deficiency at the apical site~\cite{Sawa1989}. However, because the increased oxygen content due to annealing ($\delta$) is only 0.02, the hole-doping through the introduction of oxygen is likely not the main reason for the appearance of superconductivity  within the entire range of 0.14 $\leq x \leq$ 0.28 in LESCO~\cite{Asano2019}. To understand the origin of the variation of the ground state due to annealing in the T*-type cuprate, the actual hole density ($n_{\rm h}$) should be determined. We therefore conducted an   O $K$-edge X-ray absorption spectroscopy (XAS) measurement on T*-type LESCO. 

Polycrystalline samples of AS and OA LESCO are the same as those used for previous muon spin relaxation ($\mu$SR) experiments\cite{Asano2018}. We prepared the sintered prismatic samples with a size of 1~mm $\times$ 1~mm $\times$ 2~mm for AS $x$ = 0.14 and 0.26, and OA $x$ = 0.14, 0.16, and 0.26. The samples were folded just before being placed in a vacuum chamber to obtain a clean surface. The XAS experiments were conducted at BL07LSU in SPring-8, and the absorption spectra were measured using the total fluorescence yield method at room temperature.

	\begin{figure}[t]
	\begin{center}
	\includegraphics[width=85mm]{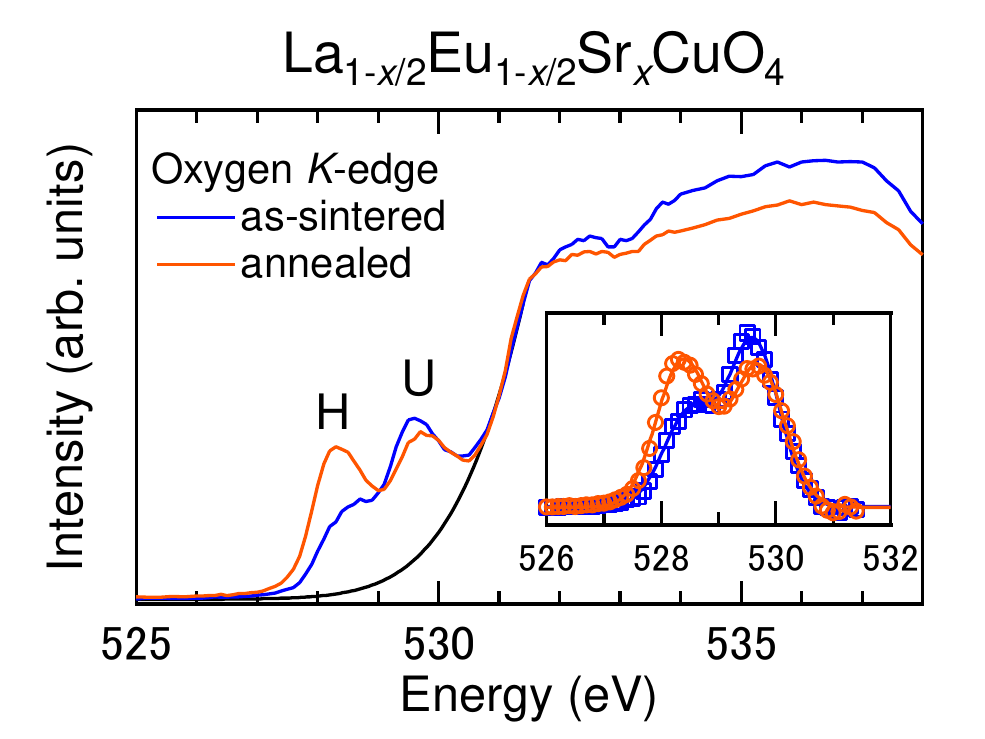}
	\caption{(Color online)~The absorption spectra of as-sintered and oxidation annealed LESCO with $x$ = 0.14.  The peaks labeled H and U correspond to a hole-induced state and a hybridized state of Cu UHB with the O 2$p$ states, respectively. The inset shows the background-subtracted spectra. }
	\label{25509Fig1}
	\end{center}
	\end{figure}

In Fig. \ref{25509Fig1}, the absorption spectra for AS and OA LESCO with $x$ = 0.14 are shown. The intensity was normalized such that the spectra at approximately 531 eV, where the slope of the absorption edge is the steepest, match well with each other. 
According to an O $K$-edge XAS study on La$_{2-x}$Sr$_x$CuO$_4$ (LSCO)\cite{Chen1991}, the peaks labeled H and U correspond to a hole-induced state and Cu upper Hubbard band hybridized with the O 2$p$ states, respectively. 
The background at below $\sim$ 531 eV, shown by the black line, was subtracted from the normalized spectra. The inset represents the subtracted spectra. As shown in the figure, the intensity of peak H, $I_{\rm H}$ (U, $I_{\rm U}$), increases (decreases) from the annealing, showing an increase in $n_{\rm h}$\cite{Chen1991}. 

\begin{figure}[tb]
	\begin{center}
	\includegraphics[width=85mm]{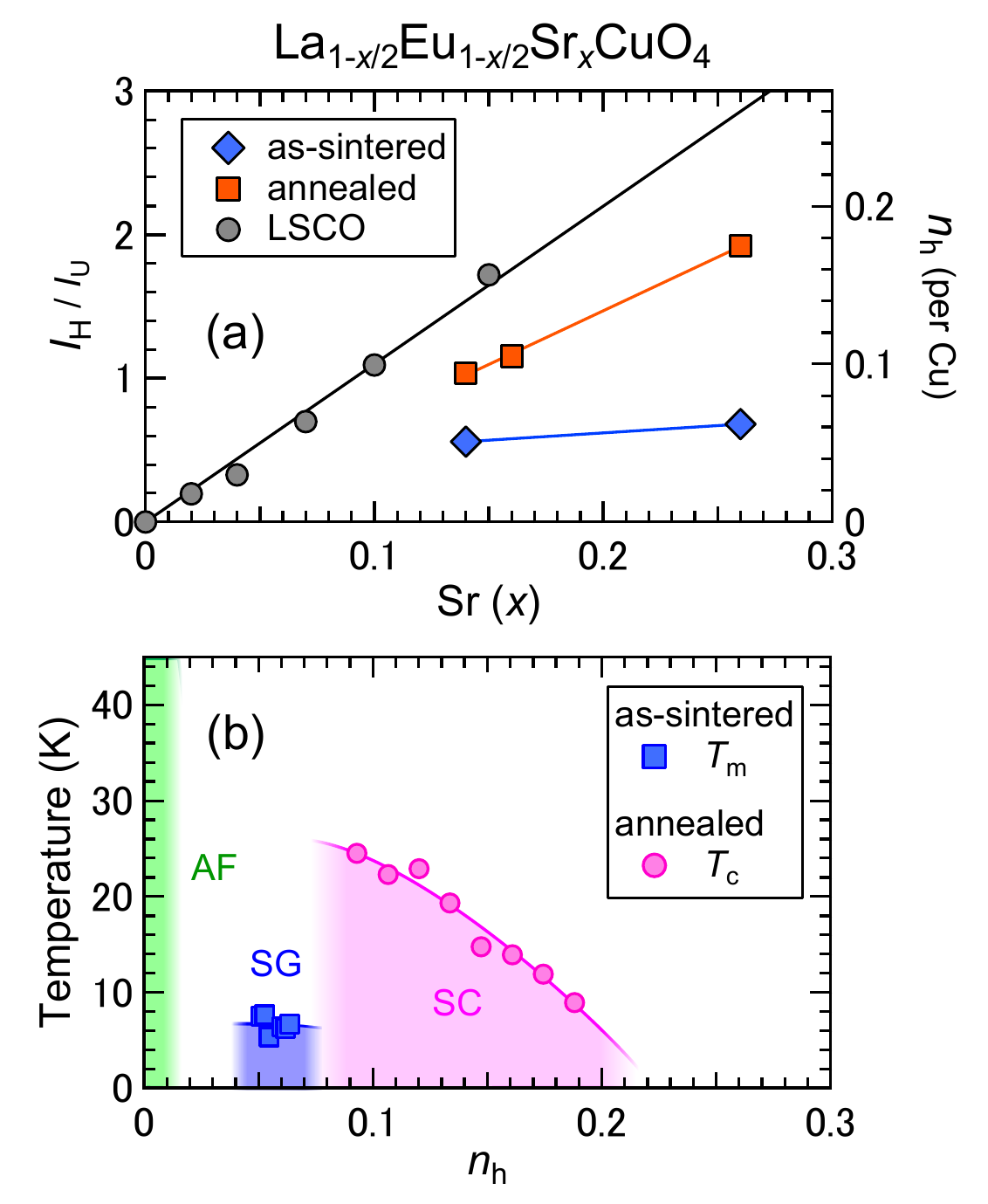}
	\caption{(Color online)~(a) Sr concentration ($x$) dependence of the peak intensity shown in Fig. \ref{25509Fig1} ($I_{\rm H}/I_{\rm U}$) and the corresponding hole density ($n_{\rm h}$) for La$_{1-x/2}$Eu$_{1-x/2}$Sr$_x$CuO$_4$. Here, $I_{\rm H}/I_{\rm U}$ for La$_{2-x}$Sr$_x$CuO$_4$ is also plotted\cite{Chen1991}. (b) Magnetic and SC transition temperatures ($T_{\rm m}$ and $T_{\rm c}$) in as-sintered and oxidation annealed La$_{1-x/2}$Eu$_{1-x/2}$Sr$_x$CuO$_4$~\cite{Asano2019} respectively as a function of the hole density evaluated from $I_{\rm H}/I_{\rm U}$ ($n_{\rm h}$). The green, blue, and pink areas represent the antiferromagnetic ordered (AF), spin-glass (SG), and superconducting (SC) states, respectively.}
	\label{25509Fig2}
	\end{center}
	\end{figure}

For LSCO, in which $n_{\rm h}$ is almost equal to $x$, a linear increase (decrease) of $I_{\rm H}$ ($I_{\rm U}$) against $x$ was reported~\cite {Chen1991}. Therefore, through a comparison with the results for LSCO, we can evaluate $n_{\rm h}$ in LESCO from the intensity ratio between $I_{\rm H}$ and $I_{\rm U}$. With this method, the intensity ratio is essential for an evaluation of $n_{\rm h}$. Therefore, the ambiguity in the normalization of the spectrum to determine $n_{\rm h}$ is eliminated. The integrated intensity for two peaks was obtained from the results of the fitting by the two Gaussian functions to the subtracted data. Figure \ref{25509Fig2} (a) shows the $x$ dependence of $I_{\rm H}/I_{\rm U}$ for LESCO. For comparison, the intensity ratio for LSCO is also plotted. 

The value on the right axis of Fig. \ref{25509Fig2} (a) represents $n_{\rm h}$, which is converted from $I_{\rm H}/I_{\rm U}$ with considering the results of LSCO. In the AS sample with $x$ = 0.14 and 0.26, $n_{\rm h}$ is $\sim$0.05 and $\sim$0.06 holes/Cu, respectively. These values are much smaller than $x$. There are two possible interpretations of the experimental results. First, holes are not sufficiently induced owing to the apical oxygen defects.  Second, most of the holes introduced by Sr substitution do not enter the CuO$_2$ plane, and holes are trapped outside the CuO$_2$ plane in the AS sample. Importantly, $n_{\rm h}$ for the OA samples is larger than that in the AS sample at each $x$, and the increase in $n_{\rm h}$ due to annealing becomes greater at a larger $x$; $n_{\rm h}$ for $x$ = 0.14 and 0.26 becomes $\sim$0.09 and $\sim$0.17 holes/Cu, which also occurs due to annealing. Thus, in contrast to the weak $x$-dependence of $n_{\rm h}$ in the AS samples, $n_{\rm h}$ in the OA samples increases upon the Sr substitution by $\sim$0.67 holes/Cu/$x$, although the absolute value is still smaller than $x$. Based on a naive image in which one introduced oxygen atom generates two holes in the system to maintain the charge neutrality, $\delta\sim$0.02 for the present samples gives the expected $n_{\rm h}$ of 0.04 holes/Cu. However, $n_{\rm h}$ increases by $\sim$0.11 holes/Cu in the $x$ = 0.26 sample, indicating that the annealing effect in LESCO has an aspect other than hole-doping. 

Figure \ref{25509Fig2} (b) shows the magnetic and SC phase diagram of LESCO in which the horizontal axis corresponds to $n_{\rm h}$. In addition, $n_{\rm h}$ for 0.14 $< x <$ 0.26 was obtained from the interpolation of $n_{\rm h}$ for 0.14 and 0.26. The SC transition temperature ($T_{\rm c}$) in the OA samples and the magnetic ordering temperature determined by $\mu$SR experiments ($T_{\rm m}$) for the AS samples are taken from the previos study\cite{Asano2018}. In the figure, we took the appearance of long-range magnetic order of below $\sim$170~K in T*-type parent La$_{1.2}$Tb$_{0.8}$CuO$_4$ into account~\cite{Lappas1994}. The spin-glass phase is located at $n_{\rm h}\sim0.06$, and the SC phase lies in the $n_{\rm h}$ region from $0.10$ to $0.20$. Although the phase diagram for $x <$ 0.14, that is, $n_{\rm h} <$ 0.05 (0.09) for the AS (OA) samples, is missing owing to the difficulty in obtaining single-phase samples, a general feature of the phase diagram against $n_{\rm h}$ is similar to that for LSCO\cite{Kastner1998}. However, we note that the results of AS and OA cannot be simply compared in the same phase diagram, because the structural disorder owing to an oxygen deficiency will affect the physical properties in the AS samples. We also note that $T_{\rm c}$ tends to increase with a decrease in $n_{\rm h}$. Thus, it would be interesting to investigate the SC property in the lightly doped region through an analogy with the superconductivity in Ce-free T'-type $RE_2$CuO$_4$~\cite{Matsumoto2009}. 

The increase in $n_{\rm h}$ due to annealing is probably related to the O$_{\rm ap}$. It is known that $T_{\rm c}$ in T*-type copper oxide increases with the hydrostatic pressure~\cite {Murayama1989}. Because the distance between the CuO$_2$ plane and the $RE_2$O$_2$ block layer is reduced under pressure, holes can effectively enter the CuO$_2$ plane by an increase in the charge transfer through O$_{\rm ap}$~\cite{Murayama1989}. Such a pressure effect suggests that the holes introduced by a Sr substitution are not fully doped into the CuO$_2$ even in the OA samples under ambient pressure, which is consistent with the present results. In the case of annealing, the lattice elongates along the c-axis, indicating an increase in  the distance between the CuO$_2$ plane and the $RE_2$O$_2$ layer. Thus, if the charge transfer increases due to annealing, it will be caused not by the change in the structural distance but by the repair of the defect of O$_{\rm ap}$. 
Because the increased amount of hole density on the CuO$_2$ plane cannot be explained only by the change in oxygen gain, the holes are trapped outside the CuO$_2$ plane in the AS sample and most of those holes are introduced into the CuO$_2$ plane due to annealing.

In summary, the increase in the hole density due to oxidation annealing in T*-type La$_{1-x/2}$Eu$_{1-x/2}$Sr$_x$CuO$_4$ was clarified through O $K$-edge X-ray absorption spectroscopy measurements. Our results show the potential contribution of the hole-doping effect on the variation of the physical properties due to annealing. 
The holes trapped outside the CuO$_2$ plane in the block layer and/or at the oxygen-deficient apical site in the AS sample are introduced into the CuO$_2$ plane due to annealing.

\section*{Acknowledgements}
The synchrotron radiation experiments were conducted at the BL07LSU of SPring-8 through a joint research in the Synchrotron Radiation Research Organization and the Institute for Solid State Physics, the University of Tokyo (Proposal No. 2019A7593). M.F. is supported by a Grant-in-Aid for Scientific Research (A) (16H02125) and K.I. is supported by a Grant-in-Aid for Scientific Research (B) (16H04004).


\end{document}